\documentclass[preprint]{vgtc}               

\ifpdf
  \pdfoutput=1\relax                   
  \usepackage{graphicx}                
  \DeclareGraphicsExtensions{.pdf,.png,.jpg,.jpeg} 
\else
  \usepackage{graphicx}                
  \DeclareGraphicsExtensions{.eps}     
\fi%

\graphicspath{{figures/}{pictures/}{images/}{./}} 

\usepackage{microtype}                 
\PassOptionsToPackage{warn}{textcomp}  
\usepackage{textcomp}                  
\usepackage{mathptmx}                  
\usepackage{times}                     
\usepackage{cite}                      
\usepackage{tabu}                      
\usepackage{booktabs}                  

\onlineid{2591}

\vgtccategory{Research}

\vgtcinsertpkg

\usepackage{soul}
\usepackage{booktabs}
\usepackage{multirow}
\usepackage{colortbl}
\usepackage{balance}
\usepackage{graphicx,subcaption}

\def\revcolor{black} 
\def\revcoloradded{black} 
\def\changed{black} 
\def\CAChanged{black} 
\renewcommand{\hl}[1]{\textcolor{\revcolor}{#1}}

\newcommand{\changebyf}[1]{\textcolor{\revcoloradded}{#1}}

\newcommand{\TVCG}[1]{\textcolor{\changed}{#1}}
\newcommand{\CAISMAR}[1]{\textcolor{\CAChanged}{#1}}

\usepackage{lipsum}

\title{The Effect of an Exergame on the Shadow Play Skill Based on Muscle Memory \TVCG{for Young Female Participants}: The Case of Forehand Drive in Table Tennis}

\author{Forouzan Farzinnejad\thanks{e-mail: forouzan.farzinnejad@hs-coburg.de}\\ %
    \parbox{1.4in}{\scriptsize \centering Coburg University of Applied Sciences and Arts} %
\and Javad Rasti\thanks{e-mail: rasti@eng.ui.ac.ir}\\ %
     \scriptsize University of Isfahan %
\and Navid Khezrian\thanks{e-mail: navid.khezrian@hs-coburg.de}\\ %
     \parbox{1.4in}{\scriptsize \centering Coburg University of Applied Sciences and Arts} %
\and Jens Grubert\thanks{e-mail: jens.grubert@hs-coburg.de}\\ %
     \parbox{1.4in}{\scriptsize \centering Coburg University of Applied Sciences and Arts}}

\teaser{
  \centering
  \includegraphics[width=0.8\linewidth]{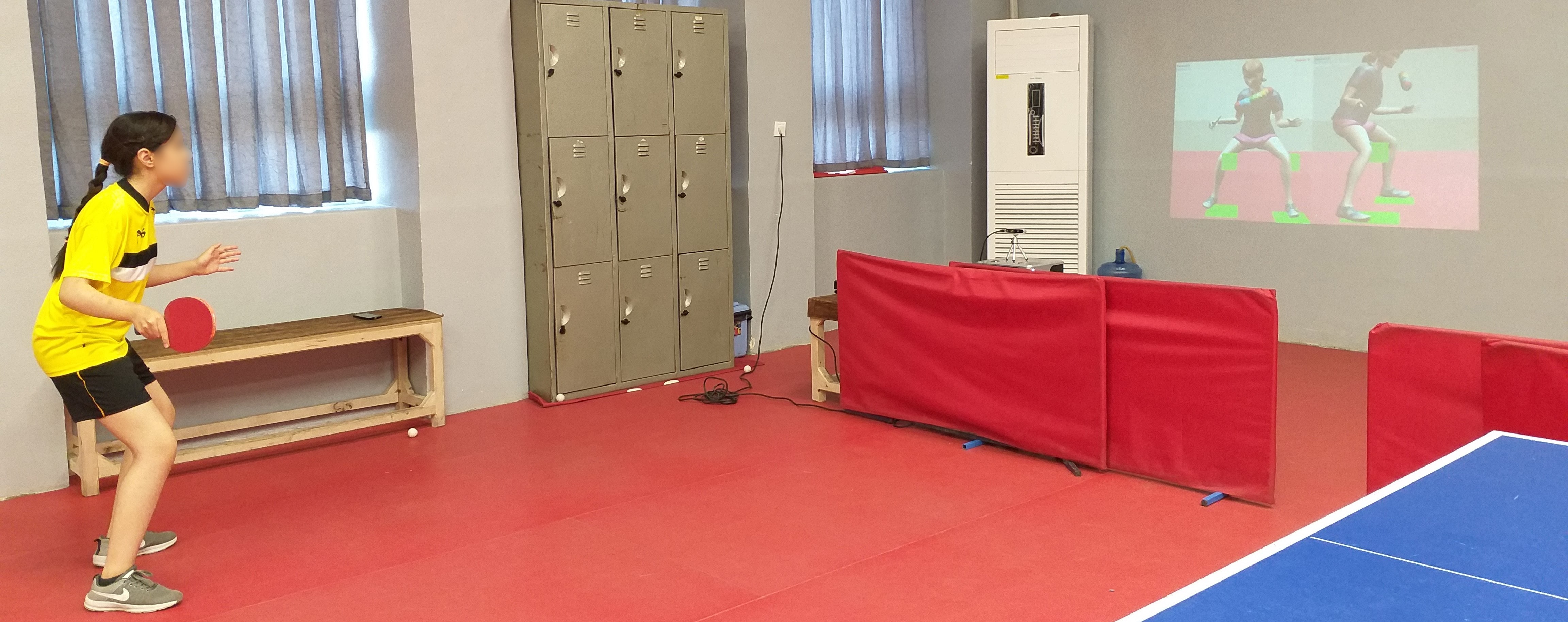}
  \caption{%
  	Practicing the forehand drive in table tennis is supported by mirroring and correcting the movements of players using a motion-controlled avatar. This results in significantly better stroke skills compared to a baseline training technique. %
  }
  \label{fig:The_proposed_system_at_work}
}

\abstract{

 Learning and practicing
 \TVCG{table tennis} with traditional methods is a long, tedious process and may even lead to the internalization of incorrect techniques if not supervised by a coach. To overcome these issues, the presented study proposes an exergame with the aim of enhancing \TVCG{young female novice players’ performance} by boosting muscle memory, making practice more interesting, and decreasing the probability of faulty training. Specifically, we propose an exergame based on skeleton tracking and a virtual avatar to support correct shadow practice to learn forehand drive technique without the presence of a coach. 
We recruited 44 schoolgirls aged between 8 and 12 years without a background in playing table tennis and divided them into control and experimental groups. We examined their stroke skills (via the Mott-Lockhart test) and the error coefficient of their forehand drives (using a ball machine) in the pre-test, post-test, and follow-up tests (10 days after the post-test). Our results showed that the experimental group had progress in the short and long term, while the control group had an improvement only in the short term. Further, the scale of improvement in the experimental group was significantly higher than in the control group. 

Given that the early stages of learning, particularly in girls children, \TVCG{are} important in the internalization of individual skills in would-be athletes, this method could support promoting correct training \TVCG{for young females}.
}

\CCScatlist{
  \CCScatTwelve{Table Tennis}{Forehand drive}{Shadow play}{Body tracking}— \CCScatTwelve{Exergame}{Avatar}{Muscle memory}{Virtual environment}{}
}

\begin{document}


\firstsection{Introduction}

\maketitle

Table tennis is one of the most popular sports in the world and a group entertainment for all age groups \cite{mueller2006table}. A common method of practice in table tennis is individual shadow play; that is, the player practices different techniques and strokes alone and without the use of a ball \cite{hodges1993table}. Due to a lack of supervision and feedback, this type of practice can lead to the internalization of problematic behaviors. This problem can be mitigated by using a coach or robot, but can impose substantial costs \cite{hodges1993table,mcafee2009table} as well as limitations concerning time and place \cite{che2014key, tabrizi2020comparative}. 

Shadow practicing different techniques such as forehand and backhand drive, cut and push, and forehand and backhand topspin is a fundamental exercise in table tennis. The most basic one of these is forehand drive which is used to strike the ball onto the right half of the court \cite{pedro2022evaluation}. Novice players must learn this technique, and, currently, the most effective method of learning is through shadow practice.
Shadow practice is the repetitive performing of a technique until it will be internalized in one’s muscle memory. Muscle memory, which is sometimes used to describe motor learning, is part of the functional memory that is responsible for memorizing a specific movement after adequate practice. When a movement is repeated frequently enough, a muscle memory record is created for it, allowing the person to do it without direct attention and concentration. Examples of muscle memory include movements that are never forgotten once learned and improved with practice, e.g. bike riding, typing, and playing a musical instrument \cite{smith2014workplace}. 
During shadow practice, players constantly follow their hand (usually before a mirror) to make sure that their hand movement is correct \cite{dubina_2020}. With the help of this exercise, the player can internalize the correct manner of the movement in their muscle memory, thereby enhancing their skills and gaining better control of the ball in real matches \cite{flores2010effectiveness}. \changebyf{It is important to note that shadow play is not only beneficial to novice players in terms of developing their stroke skills, but it can also be useful in terms of training the proper position of the racket and practicing the correct stroke technique} \cite{tabrizi2020comparative}. Therefore, shadow practice helps the player to position their muscles appropriately and hold the racket correctly so that they could concentrate on the correct method of performing a technique without being concerned with the ball. If done with sufficient attention and precision, this type of practice can be a useful training method \cite{tabrizi2020comparative}. In addition, it aids players at all levels, especially novice players \cite{flores2010effectiveness}. However, lack of a valid reference or supervisor may lead the player to wrong habits, which will not be easy to correct once they have been internalized in the muscle memory. On the other hand, as shadow practice should be repeated many times on a daily basis, it could cause boredom, which might be demotivating and reduce the quality of the exercise \cite{wu2021spinpong}.

The entertainment and education potential of computer technology, in general, \cite{vaghetti2018exergames, borna2018serious}, and the use of augmented and virtual environments, in particular, \cite{neumann2018systematic, soltani2020augmented}, can have a positive influence when teaching sports-related skills. When we participate in sports activities that involve our body and mind, endorphin is secreted in the body which makes us happy and calm. Research has confirmed the secretion of this hormone during educational computer games, thereby making the education process more interesting and the learned information more stable \cite{oliver2017gamification}. 

Further, teaching through playing is a branch of gamification that makes learning not only more meaningful but also more amusing and exciting \cite{mohammed2021motivational}. \CAISMAR{In general, gamification involves implementing game systems like competition, rewards, and user behavior analysis into contexts beyond gaming, such as work, productivity, and fitness \cite{armstrong2017evaluation}.} The learning experience enhances through gamification in all age groups, particularly if it can entertain and excite the learner and engage them actively. Due to the pleasant nature of this type of learning, the stress which is normally attributed to learning is reduced \cite{oliver2017gamification}.
In exergames \CAISMAR{(video games that need exercise activities to play \cite{boulos2013exergames,sinclair2007considerations})}, the input to a system is provided via physical movements instead of a mouse, keyboard, or joystick \cite{hayes2007incorporating}. In an exergame, instead of focusing on each other, the players focus on the screen, which helps them to concentrate on their own skills rather than on each other \cite{staiano2011exergames}. In addition to their high capacity for amusement that makes sports exercises more appealing to indolent people \cite{yim2007using}, these games precisely record the information about the athlete’s performance so that the coach can identify their points of strength and weakness and plan their training more efficiently \cite{raab2005improving}. Furthermore, exergames are a step toward e-coaching that has gained importance in the current technology-oriented world. Another advantage of these games is that they potentially influence physical, psychological, and social performance \cite{bandura1986social,bandura1999self} and increase self-esteem, motivation, and self-confidence among players for doing physical activities \cite{bandura1999self}. 

Therefore, in this study, we make use of the capacity of a virtual environment utilizing 3D body tracking to develop a shadow practice method of learning forehand drive for right-handed individuals (see \autoref{fig:the_traditional_method} showing practice in both the proposed and traditional methods). Through this, we aim to improve training outcomes over shadow practicing.

\setcounter{figure}{0}  
\begin{figure}[tbp]
  \centering
  \begin{subfigure}[b]{0.47\columnwidth}
  	\centering
  	\includegraphics[width=\textwidth]{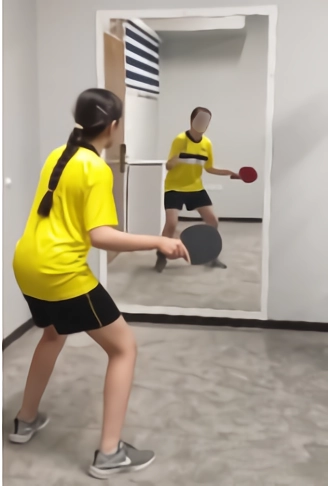}
  	\caption{The traditional method.}
  	\label{fig:ex_subfigs_a}
  \end{subfigure}%
  \hfill%
  \begin{subfigure}[b]{0.47\columnwidth}
  	\centering
  	\includegraphics[width=\textwidth]{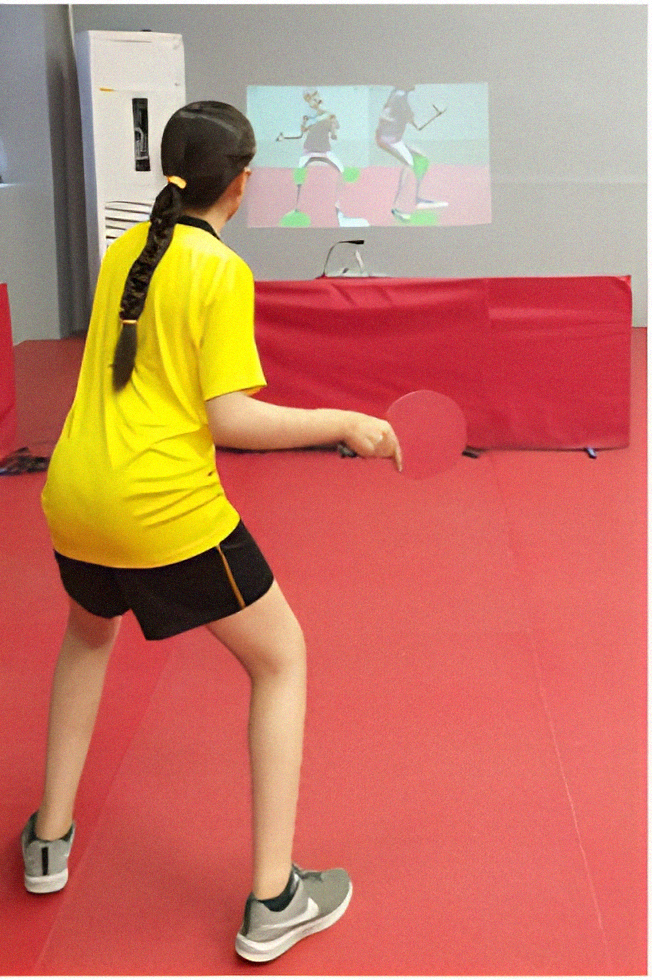}
  	\caption{The proposed method.}
  	\label{fig:the_proposed_system}
  \end{subfigure}%
  \caption{Shadow practice of forehand drive.}
  \label{fig:the_traditional_method}
\end{figure}


\hl{The main findings of our paper include that using the proposed exergame improves forehand drive skills in the short and long term \TVCG{for young females}. \CAISMAR{In our study, we directed our attention exclusively toward female participants. Our decision to exclude male participants was influenced by certain sociocultural factors that posed challenges to their recruitment within the scope of this investigation.} Through the use of the proposed exergame, it is possible to reduce the possibility of errors in forehand drive shadowing practice in the absence of a coach. Consequently, \TVCG{the young females}' forehand drive skills are improved more. Additionally, our exergame was more enthusiastically received and more motivated by the \TVCG{young female} participants than the traditional method.}
\changebyf{According to the Mott-Lockhart test, the experimental {group's} average of forehand drive strokes improved by $\mathrm{7\%}$ from the pre-test to the post-test compared with the control {group's} average. In the long-term, from pre-test to follow-up, the experimental group made more progress (\CAISMAR{$\mathrm{26\%}$}) than the control group. The average of forehand drive strokes in both groups decreased after 10 days without training, but the drop in the experimental group was \CAISMAR{$\mathrm{14\%}$} smaller than in the control group.}
\changebyf{As a result of the Stroke Error test, the mean forehand drive stroke error in the experimental group was \CAISMAR{$\mathrm{18\%}$} lower than the mean forehand drive stroke error in the control group from the pre-test to the post-test. In the long term, the mean error of the experimental group decreased by \CAISMAR{$\mathrm{22\%}$} compared with that of the control group. Despite a lack of training, mean error of forehand drive strokes in both groups increased after 10 days without training. \CAISMAR{In both groups, there was an approximately equal mean error increase.} }

\TVCG{We contribute 1) an exergame to support practicing the forehand drive technique and 2) a user study (n=44) indicating significant training benefits (as measured by the Mott-Lockhart test and Stroke Error test) over a traditional training method.}
\section{Related Work}
\label{sec:Related_Work}
\changebyf{Virtual environments have become increasingly popular in the context of sports because of their ability to provide a variety of simulations. Particularly, there is a need for improved sensorimotor skills, rather than simply using a virtual environment as a tool for strategy analysis or entertainment. Recent surveys can be found in the works of Miles et al. \cite{miles2012review}.}

\setcounter{figure}{2}  
\begin{figure}[tbp]
  \centering
  \begin{subfigure}[b]{\columnwidth}
  	\centering
  	\includegraphics[width=\textwidth]{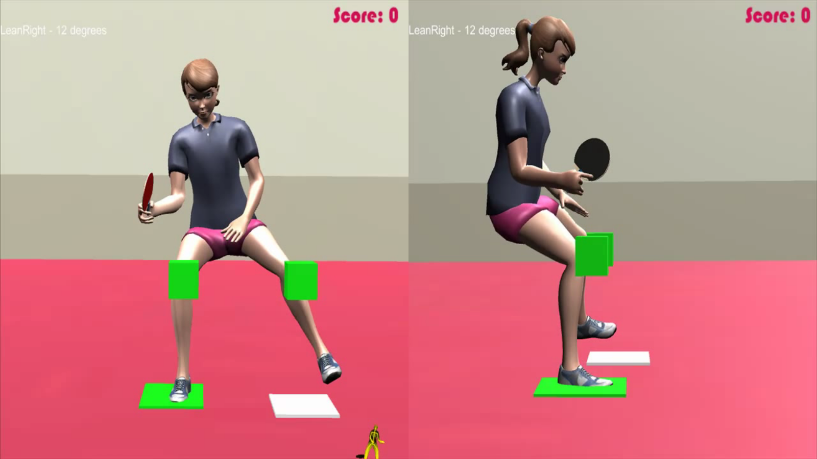}
  	\caption{Views on adjustment of the standing posture.}
  	\label{fig:adjusting_the_standing_posture1}
  \end{subfigure}%
  \\%
  \begin{subfigure}[b]{0.47\columnwidth}
  	\centering
  	\includegraphics[width=\textwidth, height=0.85\textwidth]{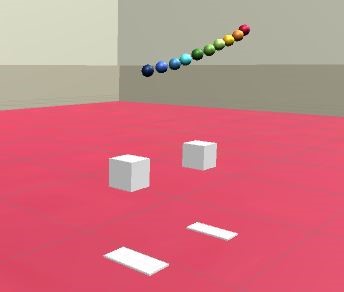}
  	\caption{The ball trajectory.}
  	\label{fig:correct-trajectory-of-ball}
  \end{subfigure}%
  \hfill%
  \begin{subfigure}[b]{0.47\columnwidth}
  	\centering
  	\includegraphics[width=\textwidth, height=0.85\textwidth]{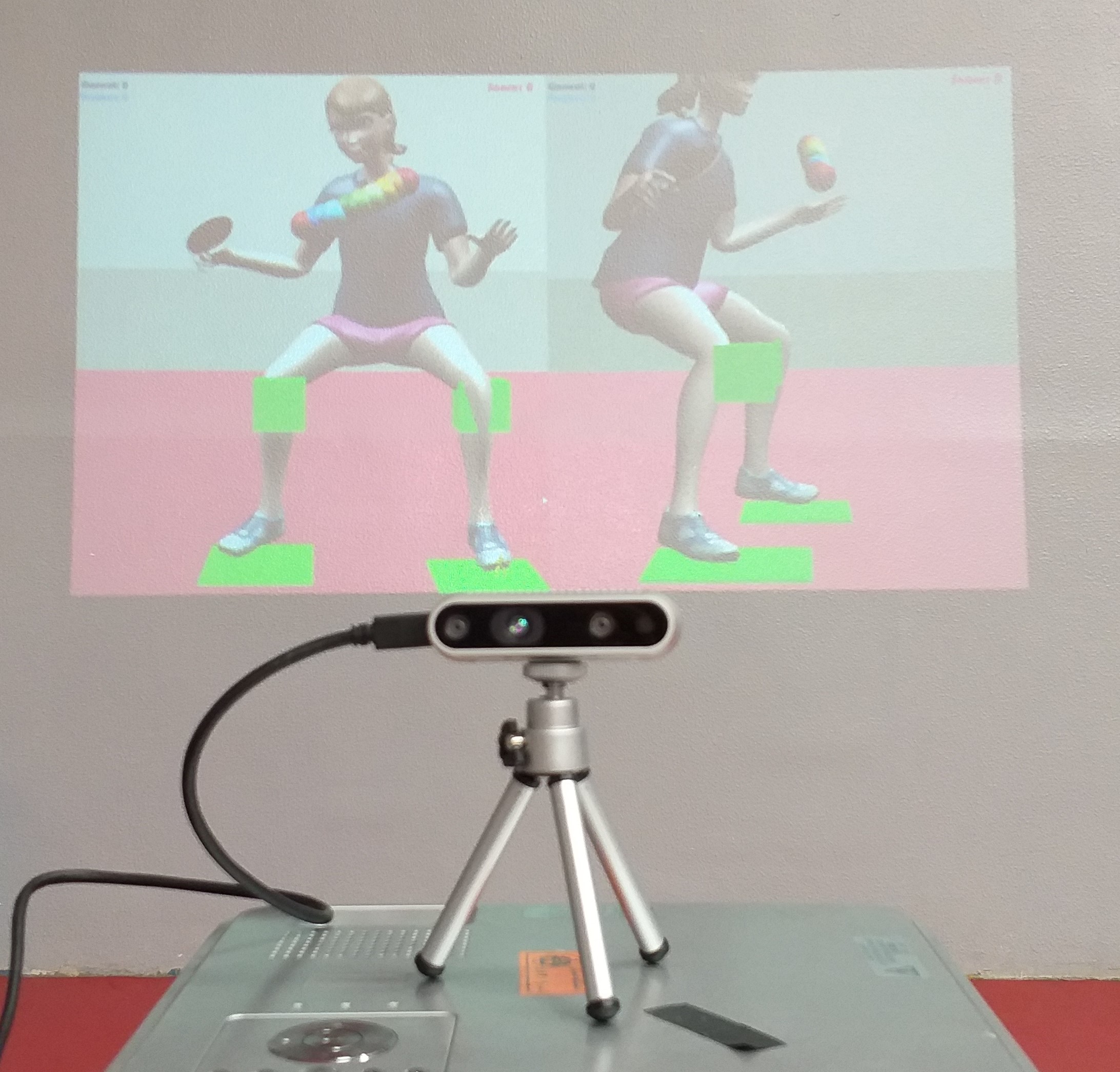}
  	\caption{Intel RealSense® depth sensor.}
  	\label{fig:RealSense}
  \end{subfigure}%
  \caption{Proposed method.}
  \label{fig:adjusting_the_standing_posture}
\end{figure}

In 2007, Bercades and Florendo \cite{florendo2007effectiveness} focused on how shadow practice could influence the learning of forehand drive. They divided 32 randomly selected subjects into control and experimental groups (each group consisting of 16 subjects). While waiting for their turn to practice with balls, the first group performed shadow practice and, at the same time, the second group practiced with a ball. The pre-test, post-test, and satisfaction scores of the participants were studied by means of adaptation and precision tests. Both groups showed significant changes in their post-test scores in comparison with their pre-test scores. However, the scores of the preservation test were only significant for the experimental group, indicating that shadow practice could be effectively used to teach the skill. Flores et al. \cite{flores2010effectiveness} investigated the correctness and precision of backhand drive among college students in 2010. The students had learned the technique through shadow practice and the results indicated the effectiveness of this practice.

Despite the fact that these two studies \cite{flores2010effectiveness,florendo2007effectiveness} examined the effectiveness of traditional shadow practice in teaching table tennis skills, traditional shadow practice has disadvantages such as potentially leading to the incorrect use of techniques if not under a coach's supervision. 

Technologies such as virtual and augmented reality along with depth sensing approaches have played a substantial role in exergames. In 2006, Brunnett et al. \cite{brunnett2006v} developed a game called V-Pong for playing table tennis in a VR environment. They investigated three groups of users with different levels of familiarity with VR and table tennis skills. They discovered that users without experience of VR environments could use the system without any training. The second group of users who were already familiar with VR technology were affected more by the quick reactions of the system. Finally, the third group who were table tennis players confirmed the almost realistic behavior of the system concerning the movement of the ball, particularly the effect of spinning. However, the fact that the table tennis players performed the techniques much faster than the other users revealed the limitations of the system. Another result of their study was the deep sense of immersion created by their system.

Mueller et al. \cite{mueller2006table} developed an online table tennis game for three (Table Tennis for Three or TTT) in 2006. In their proposed system, three persons who are at three separate tables interact with each other using real balls and rackets through augmented virtuality technology. This game provides an invaluable experience of playing and competing with geographically distant players to promote the sense of social solidarity.

In 2010, Li et al.  \cite{li2010real} developed a new real-time VR game that enabled two players to play table tennis with each other in a virtual environment. This game was equipped with a wireless tracking system that could identify the position of the player’s head and hand (racket). Also, it had two large screens for displaying the game in a 3D view. Their results indicate that the system is a suitable platform for developing other games for multiple players.

In 2012, Streuber et al. \cite{streuber2012influence} examined how error rate (task performance) and stroke speed variability (movement kinematics) are affected by different sources of visual information. According to their results, seeing the virtual player's body or paddle was crucial to preparing the stroke response. Moreover, they discovered that online arm movements are controlled in conjunction with visual information about a competitor's body. According to \cite{streuber2012influence} as visual information about the virtual player’s body and/or paddle increases task performance when the ball is invisible, we investigated shadowing practice.

\changebyf{Liu et al. \cite{liu2020virtual} developed a modular virtual reality gaming application that can be used to synthesize exercise drills for racket sports. This exergame allows users to easily adjust the objective and intensity levels of exercises. The exercise drill was optimized by using a method known as “simulated annealing” based on exercise objectives specified by the user. In the first study, heart rate readings were collected in order to evaluate the effectiveness of the developed virtual reality gaming application as an exercise tool. The second study evaluated whether participants improved their performance by using virtual reality in three different situations (no training, virtual reality training, and real-world training). Based on their findings, they concluded that virtual reality table tennis games can indeed be used for training and exercise.}

\changebyf{According to Wu et al. \cite{wu2021spinpong} in 2021, a training system was proposed for learning how to return a fast spin shot in table tennis using virtual reality cues. In SPinPong, the visual and physical sensations of playing table tennis are simulated through a VR headset and haptic feedback gloves. As part of the system, a real table tennis paddle is tracked by motion sensors and used to interact with the virtual ball. The effectiveness of SPinPong in improving table tennis skills was compared to traditional methods of training. In comparison to participants who trained with a traditional training method, SPinPong participants showed significant improvements in accuracy, speed, and spin control. According to the authors, SPinPong could be an effective tool for table tennis skills acquisition, especially for beginners.}

\begin{figure}[bp]
  \centering
  \begin{subfigure}[b]{0.47\columnwidth}
  	\centering
  	\includegraphics[width=\textwidth]{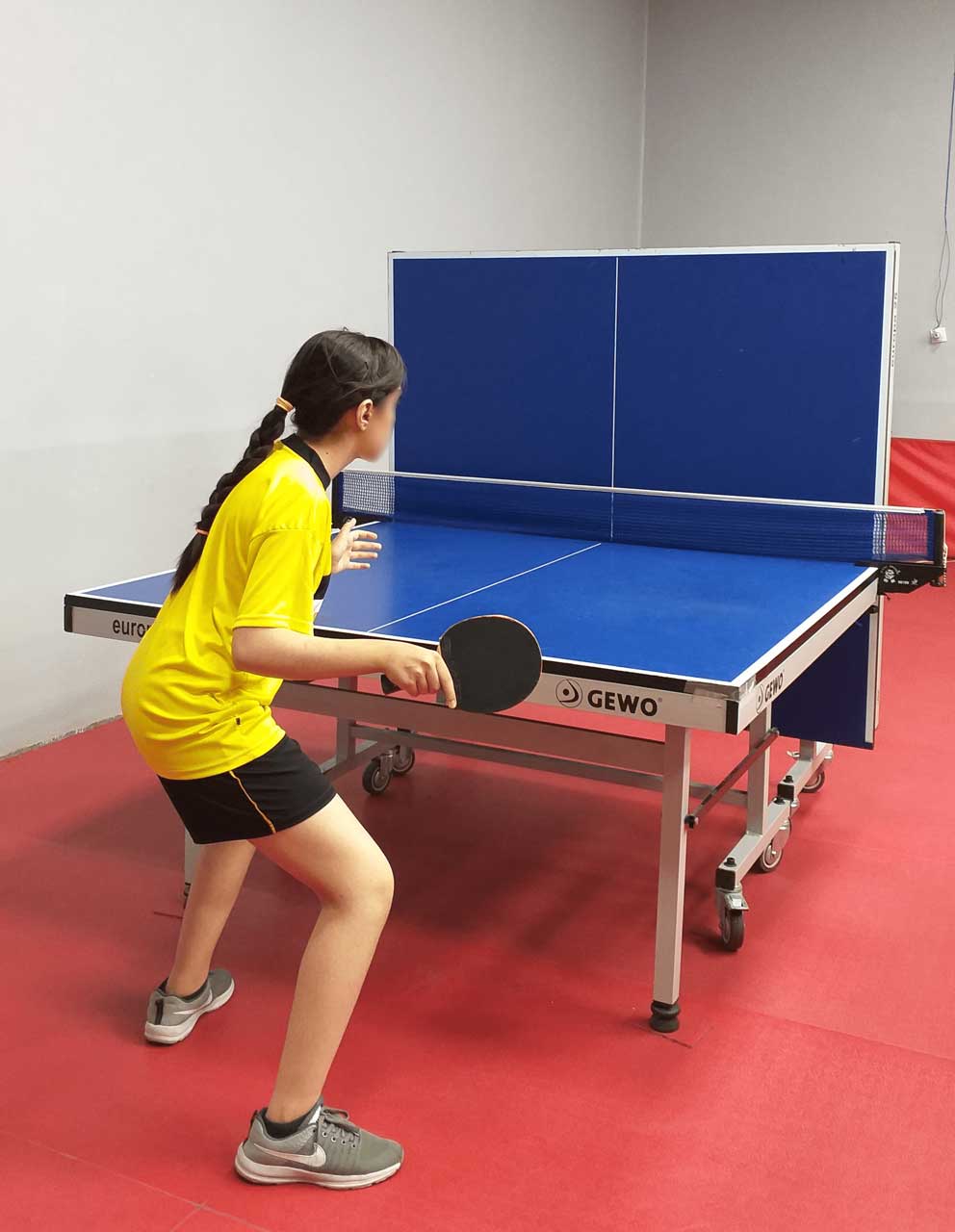}
  	\caption{Mott-Lockhart test.}
  	\label{fig:Lockhart_test}
  \end{subfigure}%
  \hfill%
  \begin{subfigure}[b]{0.47\columnwidth}
  	\centering
  	\includegraphics[width=\textwidth]{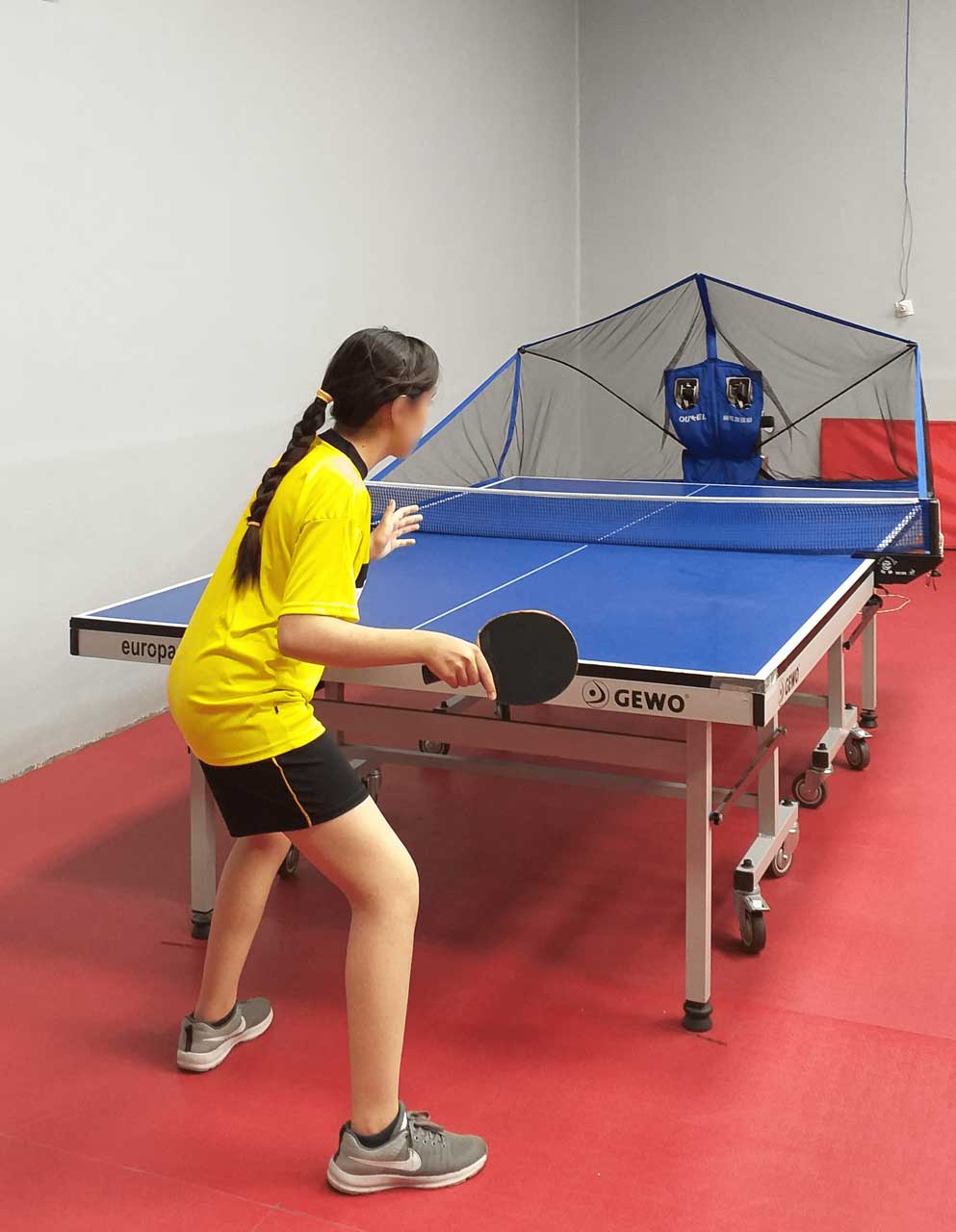}
  	\caption{Stroke Error test.}
  	\label{fig:The_ball_machine}
  \end{subfigure}%
  \caption{Evaluation tests.}
  \label{fig:test_method}
\end{figure}

 \begin{figure*}[tbp]
  \centering 
  \includegraphics[width=\textwidth]{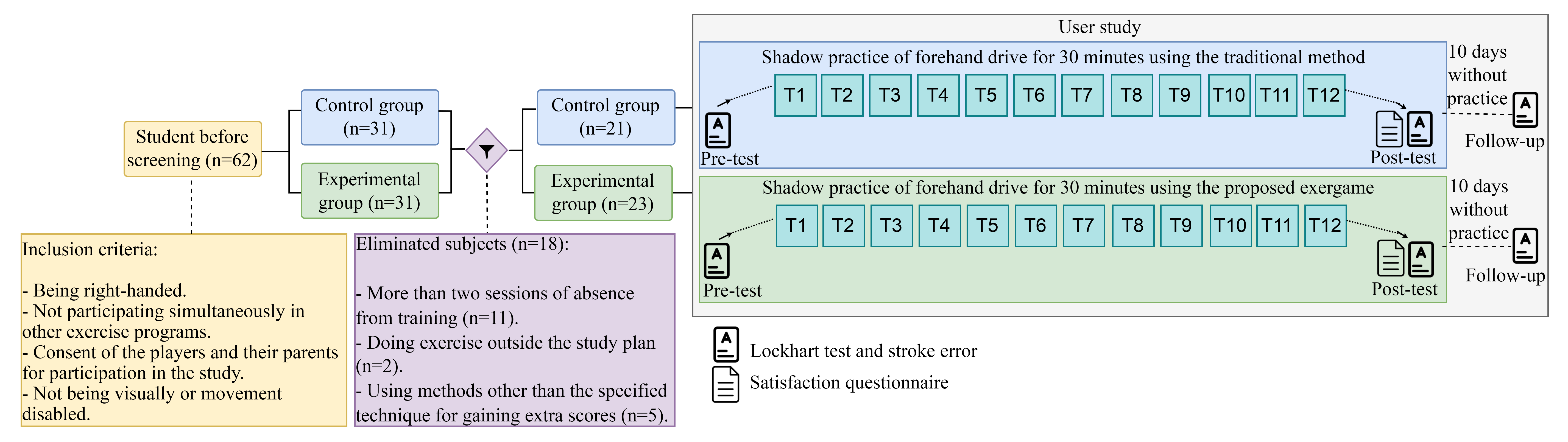}
    \caption{%
  	The treatment diagram.
  }
  \label{fig:The_treatment_diagram}
\end{figure*}

 \TVCG{Tian et al. \cite{tian2021real} in 2021 introduced an innovative football goalkeeper robot system designed to interact seamlessly with individuals and serve as a valuable tool for daily student training or entertainment purposes. This system utilizes the Intel RealSense D435 depth camera to acquire image data and the Yolov3-tiny Net to accurately localize footballs. A comparative experiment was also conducted to compare the performance of fuzzy logic control with Kalman filter algorithms. This experiment provided compelling results, demonstrating the superior efficacy of the fuzzy logic control algorithm in enhancing the overall application of the system.}
 \TVCG{In the realm of physical rehabilitation, Baldominos et. al. \cite{baldominos2015approach} proposed a novel approach for addressing rotator cuff rehabilitation. This methodology uses motion tracking technology, specifically Intel RealSense, along with virtual reality. Patients are guided through prescribed movements through a customized video game interface. In order to ensure proper execution, the system continuously monitors and verifies patients' movements. A preliminary evaluation of this approach, which has yielded encouraging results, has been conducted by experts in the field of physical rehabilitation, operating within the medical domain.  }

\begin{table*}[t]
    
    \caption{%
       \CAISMAR{ Two-way mixed ANOVA results for both tests. Gray rows show significant findings. $\mathrm{d}f_1$ = $\mathrm{d}f_{effect}$ and $\mathrm{d}f_2$ = $\mathrm{d} f_{error}$.}%
    }
    \label{tab:vis_papers}
    \scriptsize
    \centering
    \begin{tabular}{cccccccccccc}
    \hline &
    \multicolumn{5}{c}{ Mott-Lockhart Test} & & \multicolumn{5}{c}{Stroke Error Test} \\
    
    \hline &
    $\mathrm{d} f_1$ & $\mathrm{~d} f_2$ & $\mathrm{~F}$ & P  & $\eta_p^2$ & 
    &
    $\mathrm{~d} f_1$ & $\mathrm{~d} f_2$ & $\mathrm{~F}$ & $\mathrm{P}$ & $\eta_p^2$ \\

  \hline Time  & \cellcolor{lightgray}{1.59} & \cellcolor{lightgray} 65.09 & \cellcolor{lightgray} 42.92 & \cellcolor{lightgray} $<.001$ & \cellcolor{lightgray} 0.51&
    &
    \cellcolor{lightgray}1.41 & \cellcolor{lightgray}57.71 & \cellcolor{lightgray}35.88 & \cellcolor{lightgray}$<.001$ & \cellcolor{lightgray}0.47 \\

   Group  & \cellcolor{lightgray}1 & \cellcolor{lightgray}41 & \cellcolor{lightgray}6.98 & \cellcolor{lightgray}0.012 & \cellcolor{lightgray}0.14 &
    &
    1 & 41 & 0.45 & 0.508 & 0.01 \\

    Time  $\times$ Group  & \cellcolor{lightgray}1.59 & \cellcolor{lightgray}65.09 & \cellcolor{lightgray}6.09 & \cellcolor{lightgray}0.007 & \cellcolor{lightgray}0.13 &
    &
    \cellcolor{lightgray}1.41 & \cellcolor{lightgray}57.71 & \cellcolor{lightgray}3.78 & \cellcolor{lightgray}0.043 & \cellcolor{lightgray}0.08 \\

    \hline
    \end{tabular}
    \label{tab:resultsTLXSUSFlow}
\end{table*}

\changebyf{In a recent study, Oagaz et al. \cite{oagaz2021performance} investigated virtual reality (VR) training in table tennis and its potential benefits in improving performance and facilitating skill transfer. This study aims to demonstrate how VR training leads to skill transfer to real-world activity instead of comparing VR training with real training. In the study, they compared real table tennis performance between a control group with no training and an experimental group with five VR training sessions to assess skill acquisition and training transfer. According to their studies, VR training can improve table tennis performance significantly, including accuracy and speed of returns. According to the paper, VR training can be a valuable tool for improving table tennis performance and facilitating skill transfer.}

\changebyf{In order to assist beginner players in improving their playing outcomes, Oagaz et al. \cite{oagaz2022real} have created a training system that offers multi-modal training instructions and feedback on posture in real-time. Participants in the experiment were trained in the correct posture and paddling technique for forehands and backhands. The results show significant improvements in technique and ball return quality among the participants.} A VR table tennis system developed by Oagaz et al. \cite{oagaz2022real,oagaz2021performance} provides real-time feedback on posture and technique for beginner players. In contrast to our study, they did not cover skill transfer to real-world performance \cite{oagaz2022real} or the effectiveness of their tool compared to real training \cite{oagaz2022real,oagaz2021performance}.

\section{Instruments and Methods}
\label{sec:Instruments_and_Methods}

\changebyf{The purpose of our research is to develop an exergame that will help \TVCG{young female} users learn how to perform the forehand drive technique by teaching them where to position their feet, how and where to position their knees, and how to move their hands correctly. Due to the fact that forehand drive is the first and most essential skill that all beginning players must learn in order to become proficient at playing table tennis \cite{babar2021analysis}, we chose forehand drive training to include in our suggested exergame.}

In this paper, the proposed exergame was implemented by means of an Intel RealSense® D435 which is displayed in \autoref{fig:RealSense}. End-to-end latency was 30 milliseconds between making a movement and seeing it on the screen. Unity® game engine was used for the 3D implementation of the system. In this exergame, an avatar containing a human-like skeleton was designed whose movements imitate those of the player who stands in front of the depth sensor. \TVCG{Our setup effectively accommodated participants of varying heights by adjusting the height of the avatar and the corresponding ball positions to match each individual's specific height and arm length. This was made possible through the utilization of Intel RealSense® cameras, which have the capability to detect humans from backgrounds and accurately determine joint positions. The accuracy of this proposition remains consistent regardless of the player's height or size, as the software seamlessly connects the joints to a humanoid skeleton and tracks their real-time position \cite{RealSense_2022}}.

The game is displayed on a wall using a video projector so that the player watches their own performance of the techniques and receives feedback (see \autoref{fig:The_proposed_system_at_work}).

First, the way the player should position themselves for performing the forehand drive technique is presented through a text and then the correct manner of doing this technique is displayed as an animation by the avatar so that the player could become familiar with the correct forehand drive. The animation is based on the practice of one  \hl{of the top-level players and coaches in \CAISMAR{Iran}, namely \CAISMAR{Mostafa Ghazi}}, who examined the system.


\hl{For the forehand drive, the legs should be opened a little wider than the shoulder width, and if the player holds the racket with their right hand, the right leg should be slightly further back than the left leg}. 
\TVCG{We established the desired posture using four markers and positioned 10 colored balls with predetermined start and end points, following the recommended approach found in existing teaching materials \cite{muller2009table,mcafee2009table}. Furthermore, a professional athlete and experienced trainer verified the accuracy of the correct postures.}

For this reason, two virtual squares are illustrated on the ground on the screen and the player should try to put their feet on the squares. In addition, during the forehand drive, the knees should be slightly bent and the trunk leaning forward. For this purpose, two virtual cubes are shown on the screen so that the player should bend their knees and insert them into the cubes. The player’s standing style is compared against the advanced style at four points (foot soles and knees) and, on each correct adjustment, the related cube is turned green (\autoref{fig:adjusting_the_standing_posture1}). This feedback makes the player aware of the correctness of their performance. 

\TVCG{As mentioned earlier, we incorporated a mechanism to ensure the correct positioning of the user's lower body by using four white markers that would turn green when the correct position was achieved. Given the relatively limited range of movement in the upper body, we did not observe any incompatible upper-body postures throughout the study. Additionally, for the purpose of this study, we intentionally avoided visualizing fixed upper body markers. This decision was made because players typically execute a quick forward motion with their upper body just prior to striking the ball, which is crucial for accurately hitting the colored balls.}

The forehand drive technique of the above-mentioned professional table tennis player was saved as animation in \hl{the} Unity® engine and assigned to an avatar. Thus, the coordinates (x, y, z) of the points on the trajectory of the hand and the racket which were extracted were now depicted on the screen in the form of 10 colored balls to specify the correct path and scope which the novice player should follow with their racket (\autoref{fig:correct-trajectory-of-ball}).

After adjusting their style, the player should hold their arms in front of their body, bend their elbow between 90 and 110 degrees, and then move the racket in a way that it hits the 10 colored balls. Each ball has 10 scores. On each run, a score between 0 and 100 is given to the player and if the complete score is obtained, audiovisual encouragement is provided. After 10 repetitions, the player receives visual feedback: “need more effort” for 0 to 2 complete scores out of 10; “moderate performance” for 3 to 5 complete scores out of 10; “good performance” for 6 to 9 complete scores out of 10; and “excellent performance” for 10 complete scores.

\section{User Study}
\label{sec:User_Study}

We designed an experiment to evaluate our system’s efficacy for learning forehand drive in Table Tennis with shadow practice by training two groups of subjects (control and experimental groups) for 12 training sessions. \TVCG{\autoref{fig:The_treatment_diagram}} shows the experimental design. We assessed their stroke skills (using the Mott-Lockhart test \cite{mott1946table}) and their forehand drive error coefficients (using a ball machine) in pre-test, post-test, and follow-up tests (10 days after the post-test). \TVCG{Participants in both the control and experimental groups received a table tennis T-shirt as a reward after finishing the study but were not informed of this reward in advance.}

\subsection{Participants}
The participants of the study consisted of female students \CAISMAR{in Isfahan, Iran,} aged between 8 and 12 years without any experience of playing table tennis who had visited four of the sport clubs in \CAISMAR{this} city to learn this sport. \TVCG{Participants between these ages were selected as this is the age for high-potential youth players to reach the elite level\cite{faber2021developing}.} The above mentioned coach with a national coaching certificate was responsible for screening for inclusion and exclusion criteria.

The inclusion criteria were as follows: 1) being right-handed 2) not participating simultaneously in other exercise programs 3) consent of the players and their parents for participation in the study \TVCG{4) not being visually or movement disabled.}






62 subjects were selected through these criteria and randomly selected into control and experimental groups, each consisting of 31 subjects. None of the members of the two groups were aware of the activities and exercises of the other group so that it would not affect their motivation or performance.
The exclusion criteria were as follows: 1) more than two sessions of absence from training \CAISMAR{(n=11)} 2) doing exercise outside the study plan \CAISMAR{(n=2)} 3) using methods other than the specified technique for gaining extra scores \CAISMAR{(n=5)}.








After applying these criteria, 18 subjects were excluded. 23 participants remained in the experimental group and 21 in the control group. The exercise sessions were held three times a week for four weeks. 

\TVCG{In the traditional training method, following the approach described in \cite{babar2021analysis, dubina_2020}, players simulated the forehand drive stroke in front of a mirror without the use of avatars or 4-point markers. A control group shadow practiced the forehand drive for 30 minutes each session using the traditional training method, whereas the experimental group played with the same situation but instead of using the traditional method of practicing in front of the mirror, they used the proposed method.}

\subsection{Evaluation Tests}
In this study, the Mott-Lockhart test and the Stroke Error test were used to measure the players’ performance before the treatment (before the first training session), immediately after the treatment (immediately after the twelfth session), and 10 days after the end of the treatment. We compared the data from all these three sessions for statistical analysis. Also, a researcher-made questionnaire was used to assess the satisfaction levels of the experimental group participants.

\subsubsection{Mott-Lockhart Test}

The aim of the Mott-Lockhart test is to assess the ability to stroke in table tennis \cite{mott1946table}. In this test, one court of the table is folded up in a perpendicular position to the other court and used as a target board. A net is set up on the vertical court at a distance of 15 cm from the horizontal court (or the net of the table which has the same height is used) and a cardboard box containing two balls is attached to the side of the table.

The player stands in a prepared position with the racket and ball in her hands. On hearing the “ready” command, she drops the ball on the table and strikes it (with a forehand) after it rebounds so that it hits above the net on the vertical court and hits back on the horizontal court. This continues for 30 seconds and the number of strokes that hit above the net are counted as scores. The test runs three times and the best record of the player is taken as their final score. If during the test the player loses control of the ball, she can take the second ball out of the side box. When the second ball is brought into the game, the starting state of the test is repeated (see \autoref{fig:Lockhart_test}).

\subsubsection{Stroke Error Test}
The test was performed with the ball machine Oukei TW-2700-S9. The machine is shown in \autoref{fig:The_ball_machine}.

This machine throws 30 balls toward the forehand area of the player’s court which should be returned with a correct stroke onto the same area (the right side of the center line) of the machine’s side. The correctness of the stroke technique is examined by the supervisor coach. The number of errors in the 30 forehand drives is recorded for each player.

In the final session, Mott-Lockhart and Stroke Error post-tests were done on both groups. After 10 days without practice, the tests were again conducted for follow-up purposes. The treatment diagram is shown in \autoref{fig:The_treatment_diagram}.

\subsubsection{Satisfaction Questionnaire }

As one of the aims of exergame development is to use their capacity for integrating entertainment into the process of learning, measuring this feature in an exergame is of great importance. Thus, we designed a questionnaire to assess the satisfaction of the experimental group subjects with the exergame. The questionnaire consisted of five questions on a five-point Likert scale ranging from ‘very little’ (1 score) to ‘very much’ (5 scores).

\section{Results}
\label{sec:Results}

The results and analysis are based on the \CAISMAR{43} participants who had completed all the facets of the study, i.e., the participants’ satisfaction questionnaire, all training sessions, evaluation tests including the Mott-Lockhart test, and the Stroke Error test. \CAISMAR{One outlier data point was excluded.}  
We utilized a mixed within-between subjects 3x2 design. The  within-subject variable was \textsc{time} with 3 levels: \textsc{pre-test}, \textsc{post-test}, and \textsc{follow-up}. The between-subject variable was \textsc{group} with two levels: \textsc{control} group (shadow practice) and \textsc{experimental} group (exergame-mediated practice). \TVCG{Not all data were normally distributed, but normality violations do not tend to have a major impact on the robustness of the analysis from ANOVA}. Whenever the sphericity assumption was violated, \CAISMAR{we used Greenhouse-Geisser (for the Stroke Error test) and Huynh-Feldt (for the Mott-Lockhart test) corrections} but did not correct for violations of normality (c.f. \cite{blanca2023non}).
We applied Tukey-HSD correction to all post-hoc tests involving multiple comparisons.

\begin{figure}[tbp]
  \centering
  \begin{subfigure}[b]{0.49\columnwidth}
  	\centering
  	{\includegraphics[width=\textwidth]{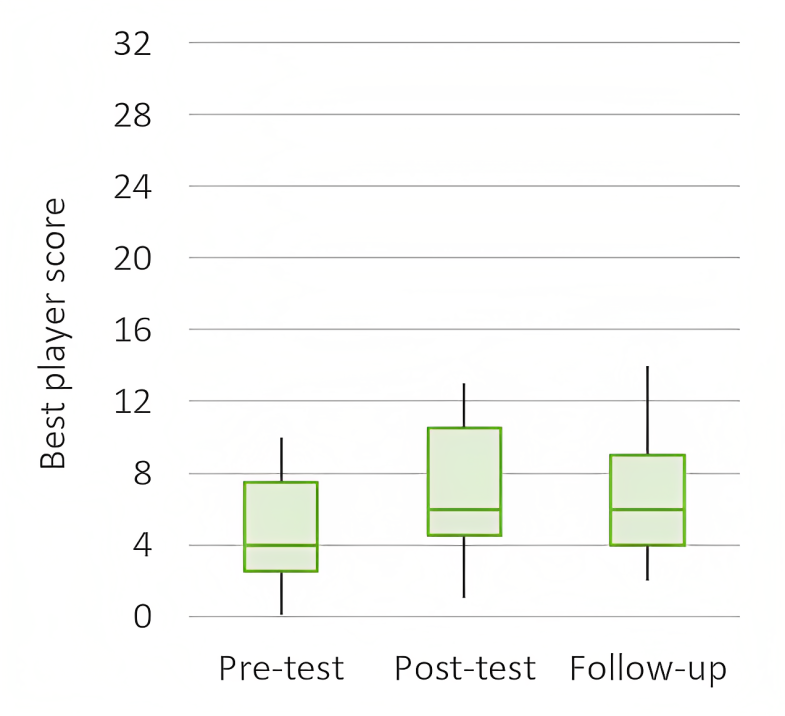}}
  	\label{fig:results_a}
         \caption{Mott-Lockhart test (Experimental group).}
  \end{subfigure}%
    \hfill%
  \begin{subfigure}[b]{0.48\columnwidth}
  	\centering
  	\includegraphics[width=\textwidth]{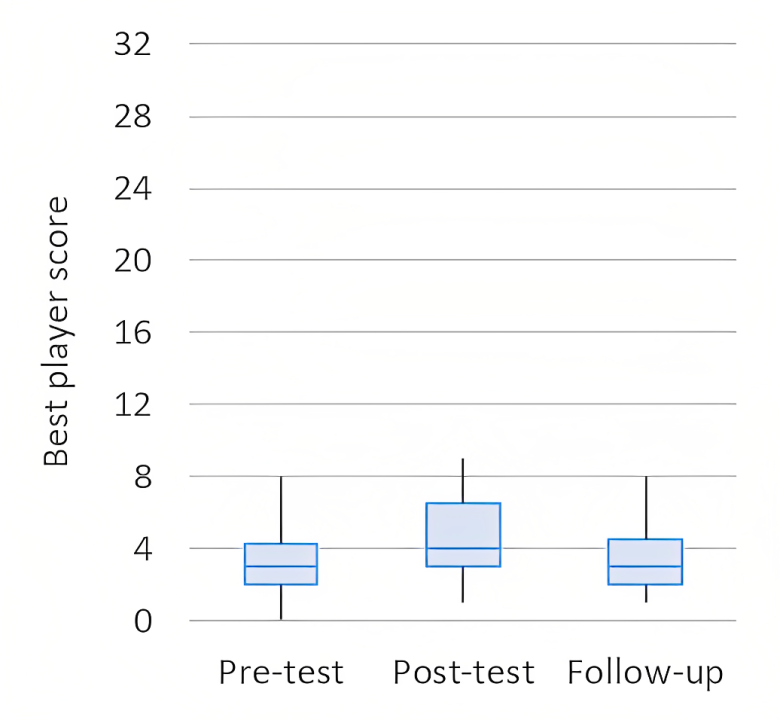}
  	\label{fig:results_b}
        \caption{Mott-Lockhart test (Control group).\newline}
  \end{subfigure}%
  \\%
  \begin{subfigure}[b]{0.49\columnwidth}
        
  	\centering
  	{\includegraphics[width=\textwidth]{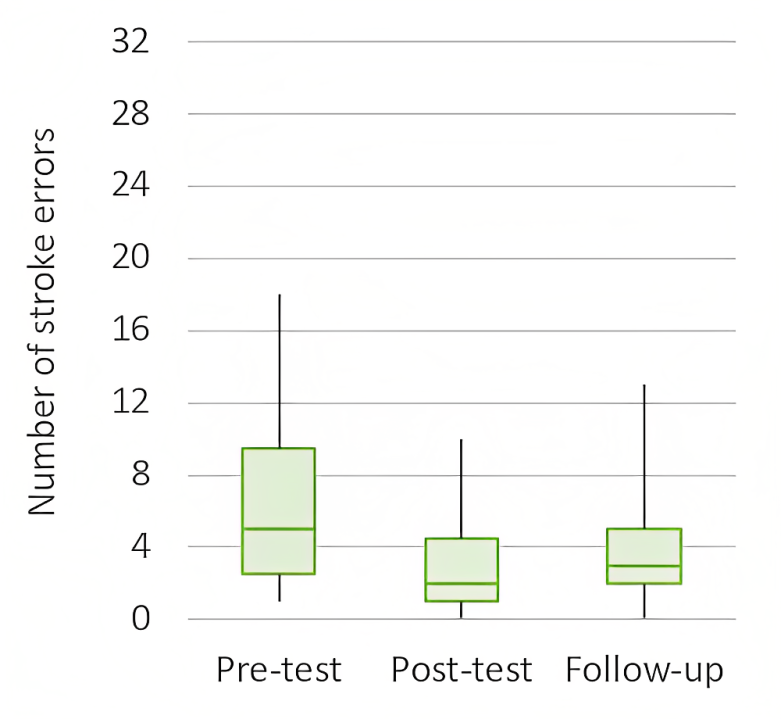}}
  	\label{fig:results_c}
        \caption{Stroke Error test (Experimental group). \newline}
  \end{subfigure}%
    \hfill%
  \begin{subfigure}[b]{0.48\columnwidth}
  	\centering
  	\includegraphics[width=\textwidth]{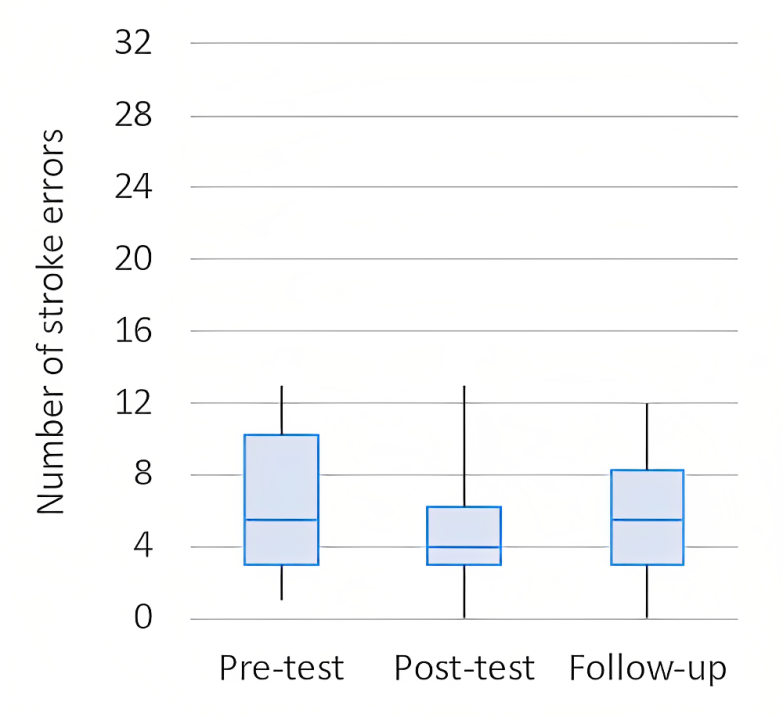}
  	\label{fig:results_d}
         \caption{Stroke Error test (Control group).\newline}
  \end{subfigure}%
  
  \caption{The descriptive results of the Mott-Lockhart and Stroke Error for each group and in each phase.}
  \label{fig:results}
\end{figure}

The Mott-Lockhart test is regarded as the standard method for evaluating stroking ability in table tennis. With the assistance of the ball machine, we also examined the number of errors that occurred during forehand drives. However, this test is not regarded as a standard method of evaluating stroking ability. Boxplots of the Mott-Lockhart and stroke error for each group are listed in \autoref{fig:results}.

\subsection{Mott-Lockhart Test}

As shown in \autoref{tab:vis_papers}, there was a significant main effect of \textsc{time} \CAISMAR{
$(\mathrm{F}(1.59,65.09)=42.92, \mathrm{p}<0.001, \mathrm{\eta_p^2}=0.51)$.}
\hl{Indicating differences between the means at each measurement \textsc{time}, ignoring the between-subjects variable}. Post-hoc comparisons with \changebyf{Tukey-HSD correction} indicate a significant difference $(\mathrm{p}<0.001)$ between the \textsc{pre-test} and \textsc{post-test}, between the \textsc{pre-test} and \textsc{follow-up}, and also between the \textsc{post-test} and \textsc{follow-up}. Also, the main effect of the \textsc{group} was significant \CAISMAR{ $(\mathrm{F}((1, 41)=6.98, \mathrm{p}=0.012, \mathrm{\eta_p^2}=0.14)$.}
 There was also a significant interaction between \textsc{time} and \textsc{group} \CAISMAR{$(\mathrm{F}(1.59, 65.09)=6.09, \mathrm{p}=0.007, \mathrm{\eta_p^2}=0.13)$.} \hl{There were differences between \textsc{group}s in how much they changed over \textsc{time}. In other words, the \textsc{experimental} and \textsc{control} groups changed at different rates.}

Post hoc comparisons using the Tukey HSD correction indicated that the mean score for the \textsc{pre-test} in the \textsc{experimental} group $(M = 4.83, SD = 3.05)$ was significantly lower than the \textsc{post-test} in the \textsc{experimental} group $(M = 7.26, SD = 3.63)$. Also, the \textsc{pre-test} in the \textsc{experimental} group $(M = 4.83, SD = 3.05)$ was significantly lower than the \textsc{follow-up} in the \textsc{experimental} group $(M = 6.61, SD = 3.27)$. However, the \textsc{post-test} in the \textsc{experimental} group $(M = 7.26, SD = 3.63)$ did not significantly differ from the \textsc{follow-up} in the \textsc{experimental} group $(M = 6.61, SD = 3.27)$. In the \textsc{control} group, the mean score for the \textsc{pre-test} \CAISMAR{$(M = 3.35, SD = 2.18)$} was significantly lower than the mean score for the \textsc{post-test} \CAISMAR{$(M = 4.80, SD = 2.61)$.} The \textsc{pre-test} in the \textsc{control} group \CAISMAR{$(M = 3.35, SD = 2.18)$} was not significantly different from the \textsc{follow-up} in the \textsc{control} group \CAISMAR{$(M = 3.70, SD = 2.39)$.} However, the \textsc{post-test} in the \textsc{control} group \CAISMAR{$(M = 4.80, SD = 2.61)$} was significantly higher than the \textsc{follow-up} in the \textsc{control} group \CAISMAR{$(M =3.70, SD = 2.39)$.}

\subsection{Stroke Error Test}
For stroke errors, we found a significant main effect of \textsc{time} \CAISMAR{$(\mathrm{F}(1.41, 57.71)=35.88, \mathrm{p}<0.001, \mathrm{\eta_p^2}=0.47)$}, see \autoref{tab:vis_papers}.
According to the pairwise comparisons for the main effect of \textsc{time} corrected using \changebyf{Tukey-HSD correction}, there was a significant difference between \textsc{pre-test} and \textsc{post-test}, between \textsc{pre-test} and \textsc{follow-up}, and also between \textsc{post-test} and \textsc{follow-up}. Also, a significant interaction was found between \textsc{time} and \textsc{group} \CAISMAR{$(\mathrm{F}(1.41, 57.71)=3.78, \mathrm{p}=0.043, \mathrm{\eta_p^2}=0.08)$.}

Based on post hoc comparisons corrected by \changebyf{Tukey-HSD correction}, the \textsc{pre-test} mean error for the \textsc{experimental} group $(M = 6.57, SD = 5.12)$ was significantly higher than the \textsc{post-test} mean error for the \textsc{experimental} group $(M = 3.61, SD = 3.38)$. Additionally, the \textsc{pre-test} of the \textsc{experimental} group $(M= 6.57, SD = 5.12)$ was significantly higher than the \textsc{follow-up} test $(M= 4.43, SD = 3.75)$. Also, the \textsc{post-test} mean error in the \textsc{experimental} group $(M = 3.61, SD = 3.38)$ was significantly lower than the \textsc{follow-up} $(M = 4.43, SD = 3.75)$ in the \textsc{experimental} group.
\textsc{Pre-test} mean error \CAISMAR{$(M =6.45, SD =4.02)$} was significantly higher than the \textsc{post-test} mean error \CAISMAR{$(M =4.70, SD =3.20)$} in the \textsc{control} group. However, the \textsc{pre-test} in the \textsc{control} group \CAISMAR{$(M =6.45, SD =4.02)$} was not significantly different from the \textsc{follow-up} in the \textsc{control} group \CAISMAR{$(M =5.75, SD =3.37)$.} the \textsc{Control} group \textsc{post-test} \CAISMAR{$(M =4.70, SD =3.20)$} was significantly lower than the \textsc{follow-up} in the \textsc{control} group \CAISMAR{$(M =5.75, SD =3.37)$.}

\subsection{Satisfaction Questionnaire}
The participants’ responses to the satisfaction questionnaire are also indicative of their great tendency to use this exergame. In response to questions 1 to 5, the percentages of the answers ‘very much’ and ‘much’ were 88\%, 88\%, 79\%, 96\%, and 96\%, respectively. It should be noted that none of the participants selected ‘a little’ or ‘hardly’ in their responses. \hl{These findings are in line with} \cite{pappas_2022} in which increased happiness of the learning process and enrichment of learning are emphasized.

\section{Discussion}
\label{sec:Discussion}

This study investigated the role of an exergame in improving forehand drive skill through shadow play \TVCG{for young females}. Shadow practice or shadow play is an exercise that helps novice players refine their skills. In most sports, including table tennis, this technique is effective for developing the skills of novice players \cite{flores2010effectiveness}. In the absence of a partner or coach, this technique can assist players in integrating the strokes they use and the grip method they use into their memory \cite{flores2010effectiveness, tabrizi2020comparative}. As a result of shadow play, players, particularly novices, are able to develop the necessary form of skill required for proper execution, as well as to form the exact stroke patterns in their minds \cite{mcafee2009table, tabrizi2020comparative}.
\changebyf{Due to potential issues of the shadow play in the traditional way \cite{wu2021spinpong}, we have developed an exergame to overcome this weakness and to increase the effectiveness of this shadow play to improve forehand drive skills \TVCG{for young females}. In order to evaluate the proposed exergame, we have used two evaluation methods, the Mott-Lockhart test and the Stroke Error test.}

According to the Mott-Lockhart test, the average of forehand drive strokes in both the post-test and follow-up of the experimental group has increased significantly compared to the pre-test. This indicates the effectiveness and progress of the players in the short and long term. As compared to the control group, the experimental group made $\mathrm{7\%}$ more progress over the short term, and \CAISMAR{$\mathrm{26\%}$} more progress over the long term. Also, there was no significant difference in the average of forehand drive strokes when comparing post-test and follow-up. The average of forehand drive strokes at both times is relatively high. This shows that 10 days without training did not affect the average of their forehand drive strokes and did not cause the players to drop substantially in performance. In the control group, there was a significant difference in the average of forehand drive strokes between the post-test and the pre-test. Therefore, there was an improvement in the short term. Even though there was a significant improvement over the short term in both control and experimental groups, the rate at which this improvement occurred in the experimental group was substantially higher ($\mathrm{7\%}$) than that of the control group in the short term. This indicates that our proposed method has been able to work more effectively than the traditional method in the ability of the forehand drive strokes. Furthermore, in the follow-up of the control group compared to the pre-test, no significant difference was seen in the number of forehand drive strokes, and as a result, there was no improvement in the long term. Also, there was a significant difference between the post-test and follow-up of the control group. The average of forehand drive strokes after 10 days of non-training is less compared to the post-test, which indicates their decline. There was a \CAISMAR{$\mathrm{14\%}$} greater drop in the control group than in the experimental group.

Based on the findings of the experimental group, in the Stroke Error test, the mean forehand drive error decreased both at the post-test and at the follow-up compared to the pre-test, suggesting that the proposed method is effective both in the short- and long-term (\CAISMAR{$\mathrm{18\%}$ and $\mathrm{22\%}$} lower error rate respectively). There was an increase in the mean error rate in forehand drive strokes in the follow-up phase when compared to the post-test. \CAISMAR{In both the control and experimental groups, the mean error increase was approximately equal}. The mean forehand drive stroke error in the control group was reduced in the post-test than it was in the pre-test. In the short term, they have prospered, but in the long run, they have not progressed. This rate of error mean reduction in forehand drive strokes in the pre-test and post-test comparison of the experimental group was more than the control group (\CAISMAR{$\mathrm{18\%}$} lower error rate). \TVCG{Therefore, the suggested approach proves to be particularly efficient when applied to young females.}
Moreover, the average of errors increased in 10 days without training in the control group. \CAISMAR{This mean error increase was $\mathrm{0.4\%}$ more in the experimental group than the control group.}

The concentration of the \TVCG{young female} players in the Mott-Lockhart test was likely higher than in the Stroke Error test, as they were required to make more forehand drives with a ball and to prevent it from falling. 
In contrast, during the Stroke Error test, each stroke was only evaluated individually, rather than a single ball being focused on for a prolonged period of time. This concentration may explain the better results in the Mott-Lockhart test.

\changebyf{Considering that the system is designed for beginners with slower movements, the accuracy and speed of the camera were sufficient for our purposes. Furthermore, RealSense® D435 is \TVCG{cost-effective}, easy to use, and \TVCG{has compact and portable design}, making it more accessible and removing barriers to adoption.}

 Our findings suggest that the positive effect of this exergame on the performance of forehand drive among \TVCG{young female} novice players is stronger than traditional methods. This effect seems to have been significantly preserved in the functional Mott-Lockhart and Stroke Error test. Our findings are in line with the work by Li et al. \cite{li2021research} that indicates the positive effect of VR technology on the technical level and training quality of athletes as well as with \cite{bedir2021effect} in which training programs containing VR-based visualization contribute to the progress of the performance of strokes. These studies show that using entertaining technologies can improve the performance of athletes and novice players.

\section{Limitations}
\CAISMAR{One limitation of this study is its exclusive concentration on young female participants, which could impact the broader applicability of the results. While this focus enabled a detailed exploration of the context surrounding young female participants, it raises concerns about the extent to which the findings can be extended to a more diverse population. The study's restriction to a single gender may not fully capture the multifaceted range of experiences and viewpoints present within the larger populace, necessitating further investigation to validate generalizability. Specifically, reported performance differences between male and female table tennis players in terms of coincidence-anticipation timing and reaction times\cite{ak2010coincidence}, grip strength\cite{carrasco2010grip}, dynamic posture control in multi-ball training\cite{gu2019effects}, and females being more susceptible to motion sickness in virtual environments \cite{peck2020mind}, and different levels of performance \cite{zagatto2016body} among table tennis players indicate that further research is needed to draw conclusions beyond the sample utilized in this study.}

\TVCG{We acknowledge the relatively high end-to-end latency of around 30 ms as the second limitation, particularly when utilizing Intel RealSense® technology. Therefore, for advanced and professional training, this latency might not be acceptable.}

\TVCG{In addition to the previously mentioned limitations, there is a third limitation to acknowledge. \CAISMAR{This study relied on self-reporting from participants regarding their adherence to the study plan and their abstention from practicing outside of it.} However, it is important to note that, in theory, participants could have provided inaccurate information or intentionally lied about their compliance. This limitation is not unique to this study but applies to any research that relies on subjective feedback.}

\section{Conclusion and Future Works}

\changebyf{Shadow practice is a common method of training muscle memory which is inherently boring due to the long time it takes to master. While shadow play can achieve desired results, professional coaches and time are required, which is often not feasible or available. Consequently, it may lead to the internalization of wrong techniques if done without the supervision of a coach. 
Accordingly, our method was developed to meet the need for an automated, intelligent training assistance solution that is available at \TVCG{cost-effective} and independent of the coach's physical presence. In this study, we designed an exergame that can accelerate the progress of a player in learning the forehand drive technique and reduce the probability of technical errors in the absence of a coach.} 

\changebyf{According to our paper, the proposed exergame improves forehand drive skills on a short and long-term basis. It is possible to reduce the risk of errors in forehand drive shadowing practice by using the proposed exergame when a coach is not present. Therefore, our exergame resulted in a greater improvement in the participants’ forehand drive skills. Moreover, the participants were more enthusiastic and motivated to participate in our exergame than with the traditional method. During the Lockhart test, the experimental group's average forehand drive strokes increased by $\mathrm{7\%}$ compared to the control group's average. The experimental group progressed \CAISMAR{$\mathrm{26\%}$} more than the control group over the long term, from the pre-test to the follow-up period. After 10 days without training, the average forehand drive strokes decreased in both groups, but the experimental group experienced a \CAISMAR{$\mathrm{14\%}$} smaller drop than the control group. The Stroke Error test indicated that the mean forehand drive stroke error in the experimental group decreased by \CAISMAR{$\mathrm{18\%}$} from the pre-test to the post-test. 
Over the long term, the experimental group demonstrated a \CAISMAR{$\mathrm{22\%}$} decrease in mean error compared to the control group. While both groups showed an increase in mean errors of forehand drive strokes after 10 days without training, the increase in the experimental group was approximately \CAISMAR{$\mathrm{0.4\%}$ higher}. On the other hand, our field evaluations are indicative of the enthusiasm and motivation of players to use this exergame for improving their skills.}

\changebyf{This particular type of training platform can be generalized and used for training purposes. \CAISMAR{In this work, we presented and studied a single optimized training system. In future work, one could study individual aspects of the training system, such as the effect of avatar representations, stroke visualization, or camera perspectives on training outcomes. In addition,} we would like to include further techniques such as backhand drive, forehand topspin, backhand topspin, and push as well as styles from a wide range of professional players. The exercises may also be tailored to meet the needs of a larger and more diverse population by using advanced approaches such as \TVCG{collecting motion data, visualizing dynamic aspects}, adaptive feedback, and machine learning. \TVCG{To enhance the external validity of future research, it is essential to include a more diverse range of participants, encompassing various genders and demographic backgrounds. This would enable a more comprehensive understanding of the phenomena under investigation and provide a more robust foundation for generalizations.} The information recorded by the computer about the players’ style of playing can help to extract their points of weakness and further customize the exercises according to their individual needs. \TVCG{Moreover, we explore the possibility of comparing our proposed approach with a different baseline, such as VR, to further evaluate its effectiveness and performance.} The treatment in this study was done over a time span of one month and the follow-up test was performed 10 days after the end of the treatment. To examine the preservation of the technique in muscle memory, we suggest that longer treatments and follow-up intervals be applied to future studies.}

\balance

\bibliographystyle{abbrv-doi}

\bibliography{template}

\end{document}